\documentclass{article}

\usepackage[T1]{fontenc} 
\usepackage{float}       
\usepackage{helvet}      
\usepackage{mathptmx}    
\usepackage{setspace}    

\usepackage{bm}
\usepackage{graphicx}
\usepackage{amssymb}
\usepackage{amsmath}
\usepackage{float}
\usepackage{xcolor}

\usepackage{pdfpages}

\begin{document}

\centerline{\bf Sign of the Gap Temperature Dependence in CsPb(Br,Cl)$_{3}$ Nanocrystals}

\centerline{\bf Determined by Cs-Rattler Mediated Electron-Phonon Coupling}



\vspace{0.5cm}

\centerline{S. Fasahat$^{1}$, N. Fiuza-Maneiro$^{2}$, B. Sch\"{a}fer$^{1}$, K. Xu$^{1}$, S. G\'{o}mez-Gra\~{n}a$^{2}$,}

\centerline{M.~I. Alonso$^{1}$, L. Polavarapu$^{2}$, and A.~R.~Go\~ni*$^{1,3}$}


\vspace{0.5cm}
\noindent
$^{1}$Institut de Ci\`encia de Materials de Barcelona, ICMAB-CSIC, Campus UAB, 08193 Bellaterra, Spain

\noindent
$^{2}$CINBIO, Universidade de Vigo, Materials Chemistry and Physics Group, Dept. of Physical Chemistry, Campus Universitario Lagoas Marcosende, 36310 Vigo, Spain





\noindent
$^{3}$ICREA, Passeig Llu\'is Companys 23, 08010 Barcelona, Spain\\
*Email: goni@icmab.es

\vspace{1cm}



\begin{abstract}

So far, the striking sign reversal in the near-ambient slope of the gap temperature dependence of colloidal CsPbCl$_3$ perovskite nanocrystals (NCs) compared to its Br counterpart, remains unresolved. Pure bromide NCs exhibit a linear gap increase with increasing temperature, to which thermal expansion and electron-phonon interaction equally contribute. In contrast, the temperature slope for the chlorine compound gap is outspoken negative. By combining temperature and pressure-dependent photoluminescence on a series of CsPb(Br$_{1-x}$Cl$_x$)$_3$ NCs, we unravel the origin of such inversion. Responsible is solely the electron-phonon interaction, undergoing a sudden change in sign and magnitude due to activation of an anomalous electron-phonon coupling mechanism linked to vibrational modes characterized by synchronous octahedral tilting and {\it Cs rattling}. This takes place in the shrunken orthorhombic NC lattice for Cl concentrations exceeding ca. 40\%. We have thus clarified a puzzling result directly impacting the optoelectronic properties of lead halide perovskite NCs.

\end{abstract}

\newpage

\section{Introduction}

Colloidal nanocrystals (NCs) of lead halide perovskites of the type APbX$_3$, with organic or inorganic A-site cation and halide substitution on the X site, exhibit very high photoluminescence (PL) quantum yields, high color purity, large band gap tunability, while being produced by low-cost, solution-processed methods \cite{ahmed20a,meixx22a,romer23a}. In this respect, fundamental knowledge about the band gap of the active material and its temperature dependence becomes mandatory for the optimization of optoelectronic devices. Under ambient conditions, metal halide perovskites, whether in bulk or nanocrystalline form, show a fairly linear gap increase with increasing temperature for either tetragonal or cubic phases. This gap temperature dependence is almost ubiquitous in halide perovskites but restricting ourselves to NCs, it can be found in MAPbI$_3$ \cite{shixx17a,rubin21a}, FAPbI$_3$ \cite{fangx17a,dirol18a}, MAPbBr$_3$ \cite{wooxx18a,sadhu19a,liuxx19a,lixxx22a}, CsPbI$_3$ \cite{saran17a,leexx17a,gauxx23a} and CsPbBr$_3$ \cite{lixxx16a,shind17a,saran17a,leexx17a,zhang19a,strand21a}. A notable exception is CsPbCl$_3$, which exhibits a sign reversal in the temperature slope, i.e. the gap decreases with increasing temperature near ambient, both for NCs \cite{saran17a} and thin films \cite{xuxxx23a}.

Recently, another hydrid perovskite with a similar temperature dependence of negative gap slope has been found, namely MHyPbBr$_3$, where MHy stands for methylhydrazinium \cite{huang22a}. This material was purposely synthesized as part of a series of lead bromides obtained by varying the A-site cation, so as to achieve a Goldschmidt tolerance factor \cite{golds26a} $t$ smaller, equal and larger than one for CsPbBr$_3$, MA$_{0.13}$EA$_{0.87}$PbBr$_3$ and MHyPbBr$_3$, respectively. Here, MA stands for methylammonium and EA for ethylammonium, while the exact composition was adjusted to obtain $t\approx 1$. 
The underlying idea \cite{huang22a} is that the stereochemical expression of the lead $6s^2$ lone pair, measured by the magnitude of the Pb off-center displacement, is predetermined by the tolerance factor. From their X-ray diffraction results and electron localization function calculations they conclude that CsPbBr$_3$ ($t<1$) is totally stereo inactive, MA$_{0.13}$EA$_{0.87}$PbBr$_3$ ($t=1$) is only dynamically stereo active and solely MHyPbBr$_3$ ($t>1$) exhibits static stereo activity. The main hypothesis is that the sign and magnitude of the electron-phonon interaction, as reflected by the slope of the gap temperature dependence, is given by the strength of the lone pair expression. Their own data support this hypothesis, for the gap temperature slope of CsPbBr$_3$ is small but positive, that of MA$_{0.13}$EA$_{0.87}$PbBr$_3$ is also positive but large and, strikingly, MHyPbBr$_3$ exhibits a clear negative slope. Regrettably, given that CsPbCl$_3$ has almost the same tolerance factor than its bromide counterpart, i.e. it should be stereo inactive, the negative temperature slope of its gap cannot be explained that way. Moreover, there are several lead-free perovskites showing positive slopes \cite{patri15a,huang21a,singh22a}, despite being much more prone to exhibit static lone pair expression. 

The effects of a temperature variation on the band structure of any semiconductor are described by the thermal expansion (TE) and electron-phonon (EP) interaction terms \cite{laute85a,gopal87a,laute87a}. The former arises from the intrinsic anharmonicity of the crystal potential, which typically leads to expansion or contraction of the crystal lattice when temperature is raised or lowered, respectively. The gap thus partly changes due to the temperature-induced volume variations. 
The renormalization of the electronic states due to lattice vibrations is essentially produced by smearing of the crystal potential and by scattering of electrons by phonons. Both effects are proportional to the Bose-Einstein phonon occupation number and so is the EP term \cite{cardo89a,shang23a}. 
Although shown for MAPbI$_3$ \cite{franc19a}, most lead halide perovskites also display a positive temperature slope of the gap which is due to similarly strong thermal expansion and electron-phonon interaction effects. Near ambient, the gap pressure coefficient of perovskites is negative, as determined by the bonding/antibonding and atomic orbital character of the band extrema \cite{franc18a}. Thus, the TE contribution to the total temperature slope is positive. It has been shown \cite{franc19a} that the EP term is usually described by a single Einstein oscillator to account for the main peak in the phonon density of states. Since its amplitude is positive, the EP contribution to the temperature renormalization of the gap is also positive. However, a recent report indicates that the incorporation of low amounts of Cs into the MAPbI$_3$ lattice leads to an {\it anomalous} electron-phonon coupling, that induces a reduction of the gap temperature slope \cite{perez23a}. The only way to account for such anomalous coupling is to introduce an additional Einstein oscillator with {\it negative} coupling constant. Here we are thus set to investigate if a similar but stronger effect could be at the origin of the negative temperature slope of the gap in CsPbCl$_3$ NCs.

In this Letter, we show results of combined temperature and pressure-dependent PL measurements, performed on a series of colloidal CsPb(Br$_{1-x}$Cl$_x$)$_3$ mixed-anion NCs to unravel the origin of the aforementioned reversal in the slope of the gap temperature dependence. The NC composition was varied continuously by ionic exchange in the primal colloidal solution. Careful analysis of the PL data allowed us to disentangle thermal expansion and electron-phonon interaction effects on gap temperature dependence. We found that, concomitant with a transition into an orthorhombic phase, occurring for Cl contents around 40\%, the electron-phonon interaction undergoes a sudden and radical change in sign and magnitude. In contrast, thermal expansion effects remain the same. Based on recent observations in mixed-cation Cs$_x$MA$_{1-x}$PbI$_3$ single crystals \cite{perez23a}, we interpret such behavior as due to the activation of an anomalous electron-phonon coupling mechanism involving Cs-cation degrees of freedom, which takes place in the shrunken orthorhombic phase stable at high Cl concentrations. The vibrational modes leading to the anomalous coupling are the so-called Cs rattlers, which have been recently invoked to understand the origin of the extremely low thermal conductivity of CsPbBr$_3$ \cite{lahns24a}.

\begin{figure}[H]
\includegraphics[width=14cm]{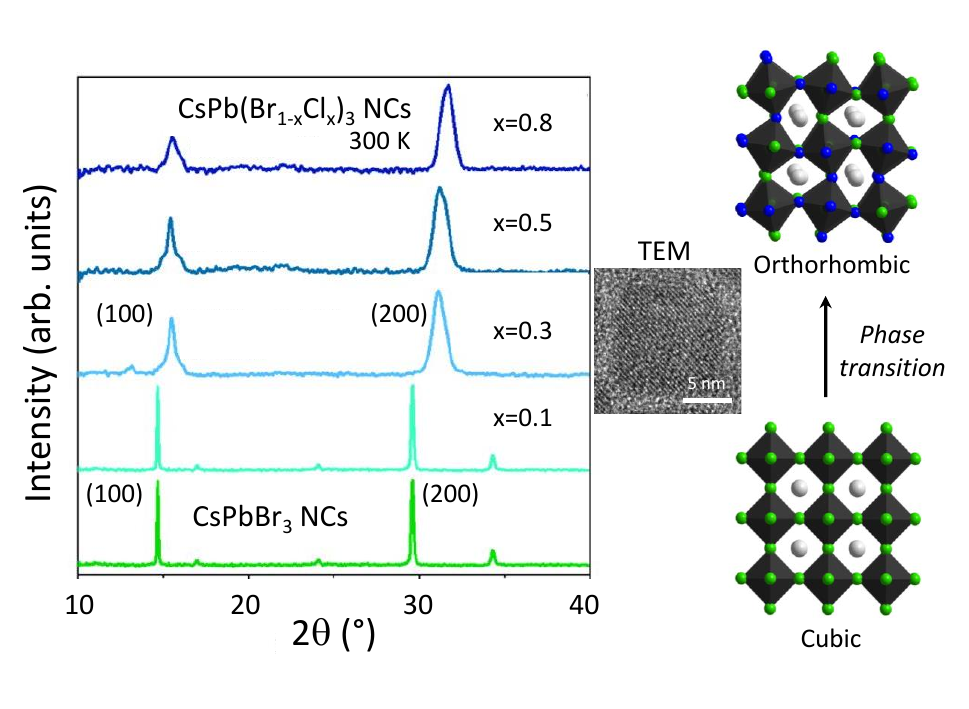}
\caption{
\label{NCs}
Room temperature X-ray diffraction patterns ($\lambda$=1.5406 {\AA}) of NC samples with similar compositions as the ones studied here. Crystal structures are also displayed (Br, Cl and Cs atoms in green, blue and grey, respectively). Center panel: High-resolution TEM micrograph of one selected NC for Cl concentration of $x$=0.75 (magnification: 400.000x).
}
\end{figure}


\section{Experimental}

First, pure CsPbBr$_3$ perovskite NCs were synthesized through ligand-assisted tip-ultrasonication, as described elsewhere \cite{tongx16a,fasah24a}. 
The prepared NCs were nearly monodisperse nanocubes with an average size ranging from 8 to 10 nm (see Fig. S1 of the S.I.). Mixed-halide CsPb(Cl,Br)$_3$ NCs were obtained by a ligand exchange strategy \cite{prote15a}, where the composition was varied continuously by performing the ionic exchange in the primal colloidal solution, until the desired emission wavelength was achieved. Figure S2 of the S.I. displays the variation of the emission color under UV illumination, resulting from the strong dependence of the NC electronic states on halide composition. In fact, the Cl content of the NCs after ionic exchange was estimated from the energy of the PL peak maximum by linear interpolation between the values for pure Br \cite{fasah24a} and pure Cl \cite{saran17a} perovskite NCs, as described in Note $\sharp$1 of the S.I. In Fig. \ref{NCs} (central panel), we show a high-resolution transmission electron micrograph (TEM) image of one selected NC of the sample with $x$=0.75, demonstrating the high crystalline quality. 

The structural characterization of the mixed-halide NCs has been carried out at room temperature by X-ray diffraction. Figure \ref{NCs} shows the two-theta scans of NC samples with different compositions but similar to the ones studied here. The assignment of the two main peaks has been performed according to the experimental and theoretical grazing-incidence wide-angle X-ray scattering (GIWAXS) results of bulk CsPbBr$_3$ \cite{hoffm23a}. The sudden shift of the diffraction peaks to higher 2$\theta$ values, occurring for Cl concentrations between 30\% and 40\% (see Fig. \ref{NCs}) is a clear indication of a structural phase transition into a phase with smaller unit cell volume. A comparison with GIWAXS \cite{hoffm23a} indicates that the sharp, non-split (100) and (200) diffraction peaks for low Cl contents are characteristic of the {\it cubic} ($\alpha$) phase. In contrast, at high Cl concentrations the shrunken phase exhibits a broader (200) peak but a relatively sharp (100) peak with two satellites. The broadening of the (200) peak is likely due to a not well-resolved splitting (see Fig. S3 in Note $\sharp$1 of the S.I.). Both changes in the X-ray diffraction line shape are indicative of a transition into the {\it orthorhombic} ($\gamma$) phase \cite{hoffm23a}. The existence of the cubic phase for low Cl contents is somewhat unexpected, for at ambient conditions the stable phase of bulk CsPbBr$_3$ is the orthorhombic $\gamma$ phase \cite{manni20a,hoffm23a}. However, high-resolution TEM studies of individual CsPbBr$_3$ NCs unraveled a size-dependent transition temperature \cite{brenn19a}. For NCs with sizes below approx. 10 nm, strain relaxation can favor the stabilization of the cubic $\alpha$ phase at room temperature, when the Cs dynamics is unfolded \cite{hoffm23a,xuxxx23b}. In fact, our Raman results obtained at ambient for the whole series of mixed-halide NCs (see Fig. S4 of the S.I. and Refs. \cite{gonix24a}) give strong support to the interpretation of the X-ray data.

\begin{figure}[H]
\includegraphics[width=5cm]{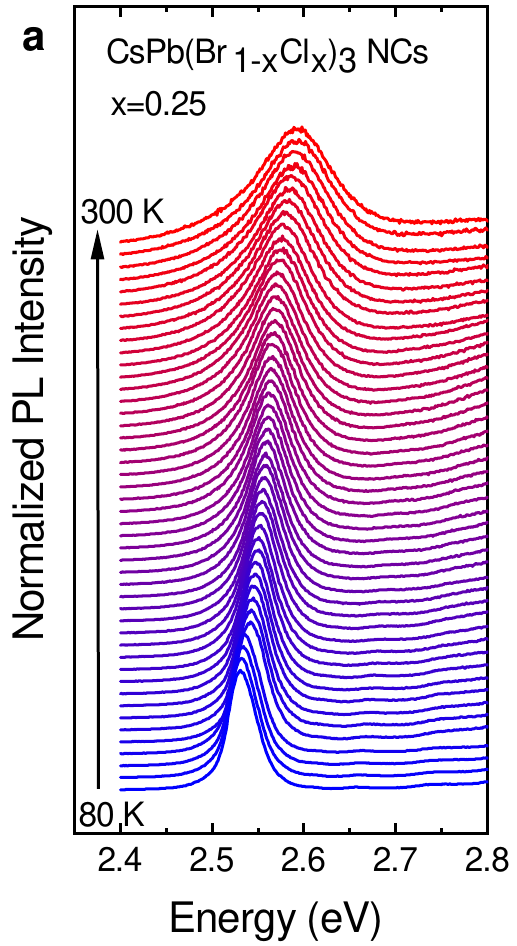}
\includegraphics[width=5cm]{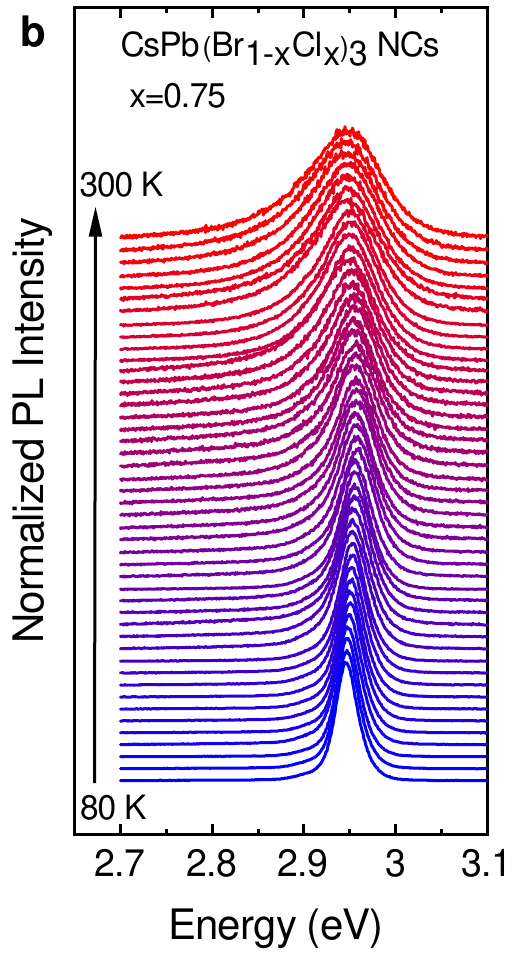}
\includegraphics[width=5.5cm]{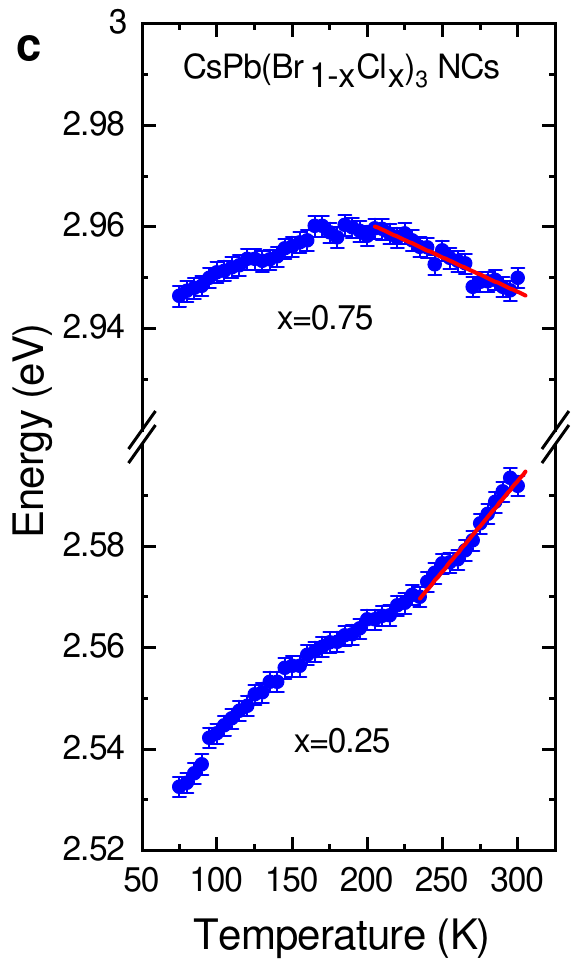}
\caption{
\label{PL-vs-T}
PL spectra of CsPb(Br$_{1-x}$Cl$_x$)$_3$ NCs with (a) $x$=0.25 and (b) $x$=0.75 measured at different temperatures between 80 and 300 K in steps of 5 K, using the 405 nm and 355 nm laser line for excitation, respectively. The spectra were normalized to their maximum intensity and shifted vertically for clarity. (c) Plot of the temperature dependence of the maximum peak position obtained from the PL spectra of (a) and (b) (blue symbols). The red lines correspond to linear fits to the data points around room temperature.
}
\end{figure}

\section{Results and Discussion}

\begin{figure}[H]
\includegraphics[width=7cm]{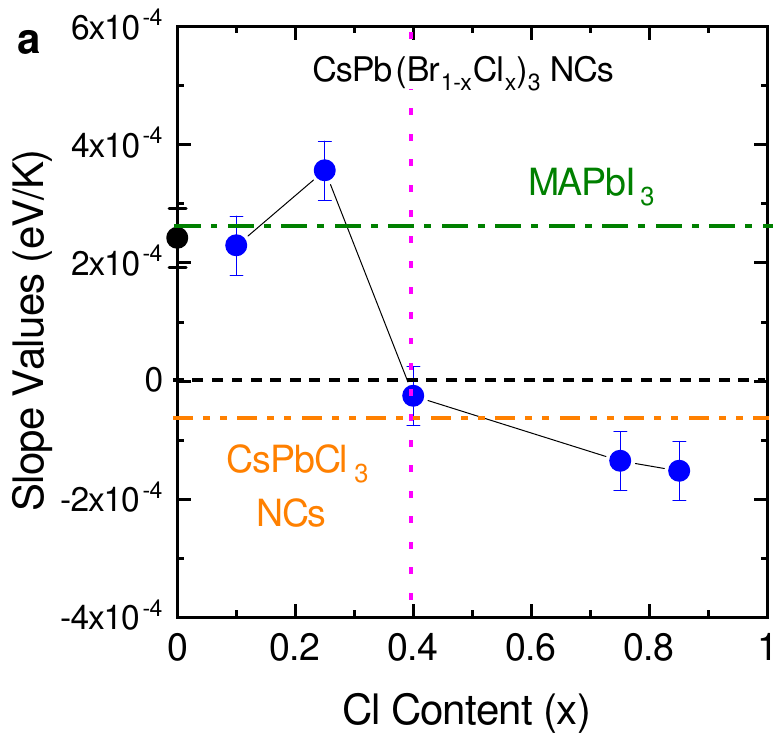}
\includegraphics[width=7.25cm]{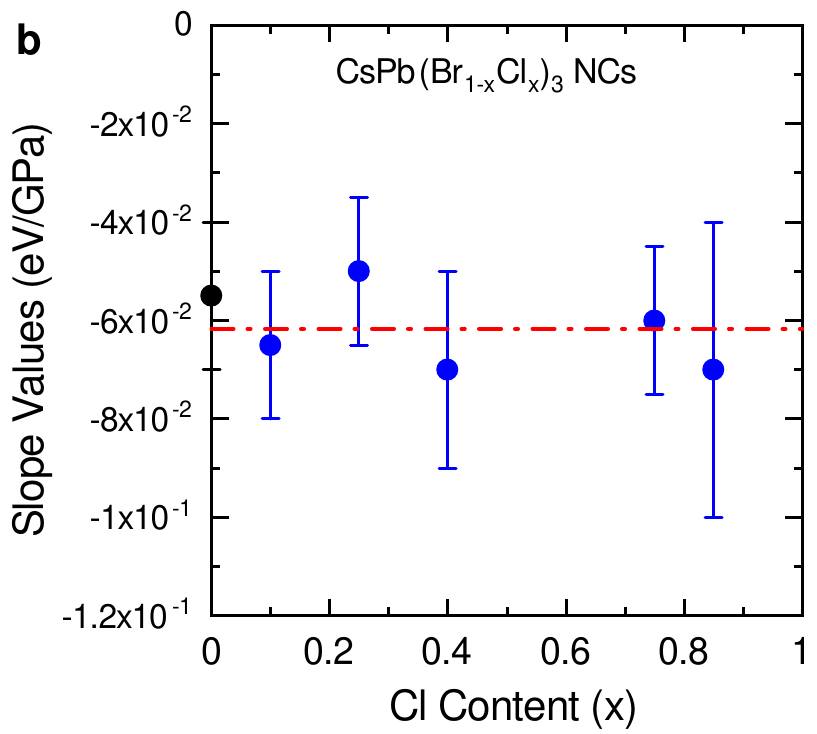}
\caption{
\label{slopes}
Slope values of the linear function of the gap energy versus (a) temperature and (b) pressure near ambient conditions of the CsPb(Br$_{1-x}$Cl$_x$)$_3$ NCs. The values for bulk MAPbI$_3$ \cite{franc19a} and CsPbCl$_3$ NCs \cite{saran17a} are indicated with dot-dashed lines. Black symbols are from Ref. \cite{fasah24a}. The red dot-dashed horizontal line in (b) corresponds to the average of the pressure slopes.
}
\end{figure}

The temperature dependent PL measurements were carried out in vacuum using a gas-flow cryostat \cite{franc19a}, whereas the high-pressure PL measurements were performed at room temperature employing a gasketed diamond anvil cell (DAC) with anhydrous propanol as pressure transmitting medium \cite{franc18a}. Figures \ref{PL-vs-T}a and \ref{PL-vs-T}b display the temperature evolution of the PL spectra of the CsPb(Br$_{1-x}$Cl$_x$)$_3$ NCs with $x$=0.25 and 0.75, respectively, from 300 K to 80 K. All spectra were normalized to its absolute maximum intensity and vertically offset for clarity. 
The values of the PL peak maximum obtained from a line-shape analysis of the PL spectra fits are plotted as a function of temperature in Fig. \ref{PL-vs-T}c \cite{franc18a,franc19a}. 
For practical purposes we consider the shift of the PL peak energy with temperature (or pressure) to be representative of the shift of the gap \cite{rubin21a,hanse24a}. The red lines in Fig. \ref{PL-vs-T}c represent the result of a linear fit to the data points around room temperature (results for other compositions are shown in Figs. S5a-c and Figs. S6a-c of the S.I.). The slopes of all gap-vs-temperature curves are plotted as a function of halide composition in Fig. \ref{slopes}a and listed in the first column of Table \ref{TE+EP}. The most striking result of this work concerns the sudden change in sign of the temperature slope, from positive, like in MAPbI$_3$ \cite{franc19a}, to negative, like for CsPbCl$_3$ NCs \cite{saran17a}, for Cl concentrations either lower or higher than 40\%, respectively. We notice that the sign turnover seems to coincide with the occurrence of the structural phase transition from cubic to orthorhombic with increasing Cl content (see Fig. \ref{NCs}).

To understand such a behavior we have first to disentangle the effects of thermal expansion and electron-phonon interaction on the gap temperature renormalization \cite{franc19a,rubin21a,perez23a}. According to Eq. (1) in Note $\sharp$2 of the S.I., the derivative of the gap over temperature contains solely the thermal expansion (TE) term and the one due to electron-phonon interaction (EP) \cite{laute85a,gopal87a,goebe98a}.
The effect on the gap due to the lattice contraction with decreasing temperature is intimately related to the response of the electronic states under hydrostatic pressure \cite{laute85a,gopal87a}:
\begin{equation}
\left[\frac{\partial E_g}{\partial T}\right]_{TE}=-\alpha_V\cdot B_0\cdot\frac{dE_g}{dP},
\label{TE}
\end{equation}
\noindent where $-\alpha_V$ is the volumetric thermal expansion coefficient, $B_0$ is the bulk modulus and $\frac{dE_g}{dP}$ is the pressure coefficient of the gap, determined here from high pressure experiments. 

The gap pressure coefficient was determined at room temperature from PL spectra recorded at different pressures, as displayed for all Cl contents in Figs. S7a-e in Note $\sharp$4 of the S.I. Figure S8 shows the variation with pressure of the gap energy, as obtained from PL line-shape fits, in the stability range of the ambient pressure phase. The slopes of the linear fits to the data points (red solid lines in Fig. S8) are plotted as a function of Cl content in Fig. \ref{slopes}b and listed in Table \ref{TE+EP}. Surprisingly, for all NC compositions the gap pressure coefficient $\frac{dE_g}{dP}$ is approx. the same in sign and magnitude, with an average value of (-60$\pm$15) meV/GPa, represented by the red dashed line in Fig. \ref{slopes}b (for a complete survey of the gap pressure and temperature coefficients of lead halide perovskites see Ref. \cite{fasah24a}). Hence, with increasing temperature thermal expansion always causes a gradual opening of the gap, independent of chlorine content. To calculate the TE term according to Eq. (\ref{TE}), we used the values of the volumetric thermal expansion coefficient $\alpha_V=(1.14\pm0.05)\times10^{-4}$K$^{-1}$ and $(1.26\pm0.05)\times10^{-4}$K$^{-1}$ from bulk CsPbBr$_3$ and CsPbCl$_3$ for low and high Cl contents, respectively \cite{nitsc02a}, and the same bulk modulus of $B_0=(21\pm5)$ GPa from CsPbBr$_3$ \cite{ezzel21a} for all compositions.  Table \ref{TE+EP} contains the so-computed TE term values. Consequently with the steadiness of the gap pressure coefficients, the TE contributions are quite independent of the halide composition, being all positive and similar in magnitude, within experimental uncertainty. This implies that the sign reversal comes from changes in the electron-phonon interaction.

\begin{table}[H]
\caption{Measured linear gap temperature ($\frac{dE_g}{dT}$) and pressure ($\frac{dE_g}{dP}$) coefficients for $x$=0, 0.1, 0.25, 0.4, 0.75 and 0.85. Also listed are the values of the TE term ($\left[\frac{\partial E_g}{\partial T}\right]_{TE}$), computed after Eq. (\ref{TE}), and the results from least-squares fits to the first derivative data points corresponding to the EP terms ($\left[\frac{\partial E_g}{\partial T}\right]_{EP}$) and the total temperature renormalization ($\left[\frac{dE_g}{dT}\right]_{TE+EP}$). Numbers in parentheses are error bars. Asterisk indicates data are from Ref. \cite{fasah24a}.}
\vspace{0.5 cm}
\begin{tabular}{| c | cc | cc | c |}
\hline
 $x$ &~ $\frac{dE_g}{dT}$ ~~&~ $\frac{dE_g}{dP}$ ~~&~ $\left[\frac{\partial E_g}{\partial T}\right]_{TE}$ ~&~ $\left[\frac{\partial E_g}{\partial T}\right]_{EP}$ ~&~ $\left[\frac{dE_g}{dT}\right]_{TE+EP}$ \\
 &~ ($10^{-4}$~eV/K)~~&~ (eV/GPa)~~&~ ($10^{-4}$~eV/K)~&~ ($10^{-4}$~eV/K)~&~ ($10^{-4}$~eV/K \\
\hline
\hline
0$^{*}$ & 2.3(5) & -0.055(15) & 1.3(4) & 0.9(2) & 2.2(3)\\

\hline
0.10 & 2.2(5) & -0.065(15) & 1.5(5) &0.8(2) & 2.3(3)\\

\hline
0.25 & 3.1(2) & -0.050(15) & 1.2(5) & 1.8(2) & 3.0(3)\\

\hline
0.40 & 0.3(5) & -0.070(20) & 1.9(4) & 1.0 & 0.3(3)\\
 &  &  &  & -2.6(2) & \\

\hline
0.75 & -1.4(5) & -0.060(15) & 1.6(4) & 1.0 & -1.4(3)\\
 &  &  &  & -4.0(2) & \\

\hline
0.85 & -1.5(5) & -0.070(30) & 1.9(7) & 1.0 & -1.2(3)\\
 &  &  &  & -4.1(2) & \\

\hline
\end{tabular}
\vspace{0.5 cm}
\label{TE+EP}
\end{table}


Regarding the contributions to the gap temperature renormalization stemming from electron-phonon interactions, the most important ones arise from peaks in the phonon density of states (DOS) \cite{gopal87a}. This is at the origin of the Einstein-oscillator model \cite{goebe98a,serra02a,bhosa12a}, which approximates the different contributions to the EP term by effective oscillators with effective amplitude $A_i$ and phonon eigen-frequency $\omega_i$, inferred from the peaks in the phonon DOS (see Eq. (6) in Note $\sharp$5 of the S.I.). The EP correction to the gap is obtained by calculating analytically the first derivative of the Bose-Einstein occupation number with respect to temperature as:

\begin{equation}
\left[\frac{\partial E_g}{\partial T}\right]_{EP} = \sum_i \frac{A_i}{4T}\cdot\frac{\hslash\omega_i}{k_BT}\cdot\frac{1}{sinh^2\left(\frac{\hslash\omega_i}{2k_BT}\right)}. 
\label{EP}
\end{equation}
\noindent Here $n_B(\omega_i,T)=\left(e^{\beta\hslash\omega_{i}}-1\right)^{-1}$ with $\beta=\frac{1}{k_BT}$ stands for the Bose-Einstein phonon occupation factor of the $i^{th}$ oscillator.

\begin{figure}[H]
\includegraphics[width=7cm]{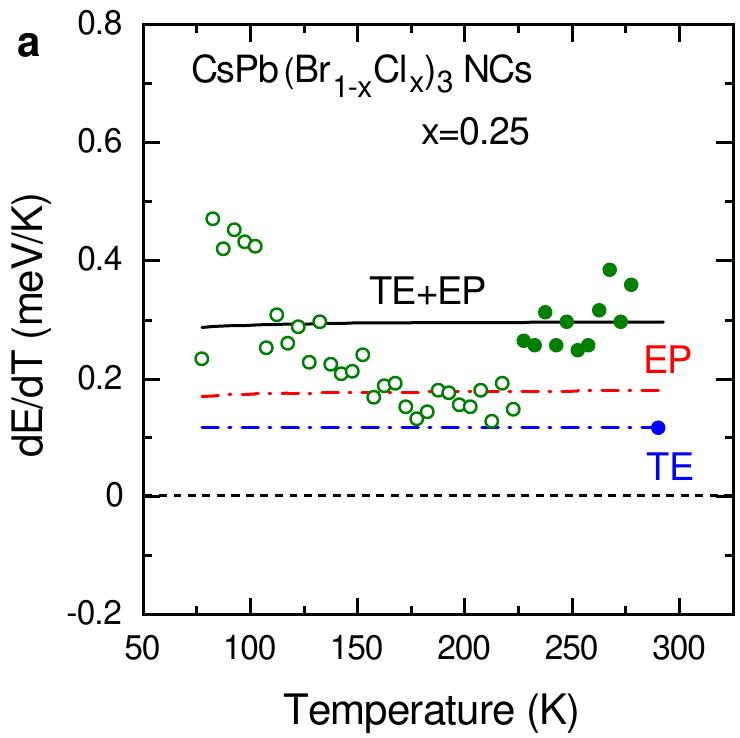}
\includegraphics[width=7cm]{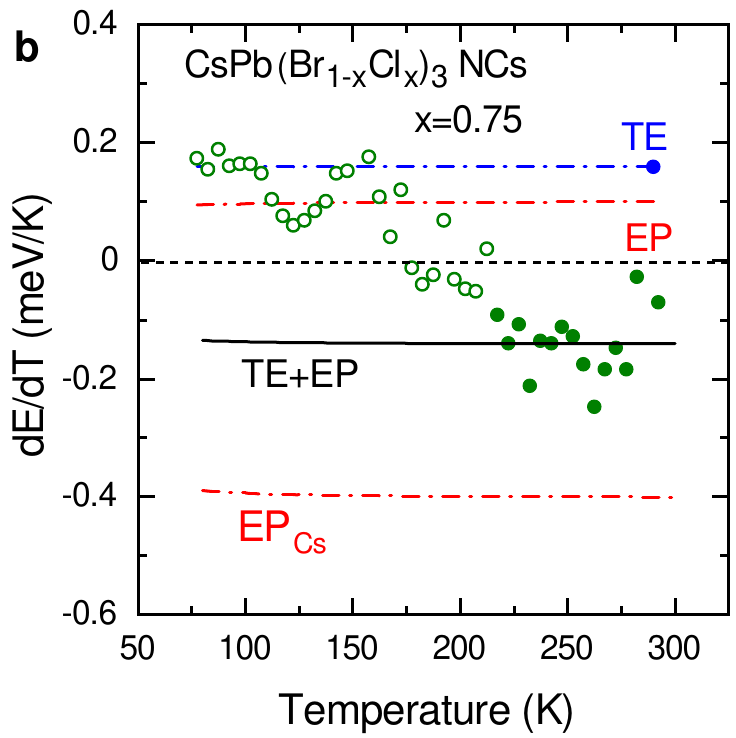}
\caption{
\label{dE-dT}
The first derivative of the gap energy with respect to temperature (dark-green symbols), numerically calculated from the smoothed data of Fig. \ref{PL-vs-T}c for CsPb(Br$_{1-x}$Cl$_x$)$_3$ NCs with (a) $x$=0.25 and (b) $x$=0.75. The solid black lines represent a fit to \emph{only} the closed data points, using the sum of thermal expansion (TE, blue dot-dashed curve and symbol) and electron-phonon interaction (EP, red dot-dashed curve). In (b), the curve labeled EP$_{Cs}$ represents the additional Einstein oscillator introduced to account for the anomalous EP coupling. See text for details.
}
\end{figure}

To evaluate the EP term we thus have to calculate numerically the first derivative of the gap energies over temperature on a point-by-point basis. To avoid unwanted amplification of the scatter of the first-derivative data points, we slightly smoothed the data sets prior to derivation using a five-points average method (see Figs. S9a-e of Note $\sharp$5 of the S.I.). The dark-green symbols in Figs. \ref{dE-dT}a,b correspond to the first derivative data sets for a Cl content of $x$=0.25 and 0.75; two representative compositions below and above the turnover composition, respectively (see Figs. S10a-c of the S.I. for the resting compositions). To obtain the EP term (or terms) by fitting, \emph{only} the data points represented by closed symbols in Figs. \ref{dE-dT}a,b and Figs. S10a-c were considered. These are the data points within the range of linearity, coinciding with the temperature range of the red lines in Fig. \ref{PL-vs-T}c and Figs. S9a-e (See Note $\sharp$3 of the S.I. on the importance of linearity). The blue dot in Figs. \ref{dE-dT}a,b corresponds to the TE term resulting from Eq. (\ref{TE}) and tabulated in Table \ref{TE+EP}. The blue dot-dashed line indicates that the TE term is temperature independent, at least in the linearity range. The contribution from electron-phonon interaction is calculated using the function of Eq. (\ref{EP}). As for the archetypal perovskite MAPbI$_3$ \cite{franc19a}, for Cl contents lower than the turnover composition at ca. 40\% a single Einstein oscillator is sufficient to account for the EP term. For MAPbI$_3$ the oscillator has a positive amplitude and a frequency of 6 meV \cite{franc19a}. An inspection of the phonon DOS for the three methylammonium lead halide (Cl, Br, and I) compounds \cite{leguy16a} indicates that the oscillator frequency of 6 meV lies slightly above the first well-defined peak or band in the DOS, corresponding to acoustical and low-frequency optical modes (see Note $\sharp$6). Since these are all phonons of the inorganic cage, we can safely assume that a similar oscillator will account for the EP term in the mixed-halide NCs as well. We have thus fixed the oscillator frequency, leaving only its amplitude as adjustable parameter. The function of Eq. (\ref{EP}) together with the constant contribution from TE were fitted to the data points (only closed symbols) of Figs. \ref{dE-dT}a,b and Figs. S10a-c. The resulting amplitudes are listed in Table S2 (S.I.). The solid black curves and the dashed red curves represent the resulting total rate of gap renormalization per Kelvin and the EP contributions to it, respectively.

For Cl contents higher than the turnover concentration the situation with the EP term changes dramatically. Due to the fact that the TE term and the EP contribution from the coupling to the inorganic cage phonons are both positive, the only way to attain a negative temperature derivative is to introduce an additional Einstein oscillator with negative amplitude. We thus used for the EP term two Einstein oscillators: One with both fixed frequency and (positive) amplitude as for the single oscillator case, and one with adjustable (negative) amplitude but frequency fixed to a value of 4 meV, the frequency of the Cs rattle modes \cite{lahns24a} (see discussion below). 
The red dot-dashed curves labeled EP$_{Cs}$ in Fig. \ref{dE-dT}b and Figs. S10b,c represent the contribution of the additional Einstein oscillator to the EP term. 

The results of the least-squares fits are shown in Table S2 of Note $\sharp$5 in the S.I. Strikingly, the amplitude of the additional oscillators with fixed (4 meV) rattler frequency ranges from -10 to -20 meV, thus overcompensating thermal expansion and the normal EP term. 
We recall that the most intense Raman modes have frequencies below 100 cm$^{-1}$ (ca. 12.5 meV) \cite{leguy16a,yaffe17a}, thus, a frequency of 4 meV matches well the vibrational spectrum of the inorganic cage. Finally, in Table \ref{TE+EP} the values computed at room temperature of the TE term, the two contributions to the EP term and the total temperature renormalization of the gap (TE+$\sum_i$EP$_i$) are listed for all studied compositions. As a consistency check, we note the excellent agreement between the experimental temperature slopes of the gap and the values obtained from the fits to the first derivative data points (first and last column of Table \ref{TE+EP}, respectively).

About the possible origin of the anomalous EP coupling term, we provide an explanation supported by similar observations made by the
incorporation of small amounts of Cs in two Cs$_x$MA$_{1-x}$PbI$_3$ single crystals (x = 0.05 and 0.1) \cite{perez23a}. In the Cs-containing samples and above a certain onset temperature of around 260 K, the slope of the linear temperature dependence of the gap reduces to about half the value of MAPbI$_3$. 
Such behavior was explained by the introduction of an extra Einstein oscillator with negative amplitude associated to the appearance of an additional electron-phonon coupling mechanism. The latter was attributed to dynamic tilting of the PbI$_6$ octahedrons in {\it synchrony} with the translational dynamics of the Cs cations between equivalent potential minima of the cage voids; a dynamics unfolded above the onset temperature. For the CsPb(Br,Cl)$_3$ NCs we will propose a similar interpretation but with nuances, accounting for differences in crystal structure and halide ionic radius.

Recently, the constant-energy surfaces of the atomic potential energy were calculated by density functional theory as a function of octahedral tilting angle for CsPbI$_3$ and CsSnI$_3$ \cite{klarb19a,patri15a}. In the orthorhombic phase, as for the NCs with high Cl contents, these potential energy surfaces exhibit a multi-well landscape \cite{klarb19a,patri15a}. 
The key point is that the shrunken lattice of the orthorhombic phase leaves very little space for the Cs cations to move inside the cage voids. Thus, even though they are locked, the Cs cations oscillate around the minimum of the atomic potential. These oscillations are the aforementioned {\it Cs rattlers}, to which the extremely low thermal conductivity of CsPbBr$_3$ have been attributed \cite{lahns24a}. The molecular dynamics (MD) calculations of the lattice thermal conductivity show that its strong reduction arises from lattice anharmonicities resulting from the effect of dynamic disorder introduced by Cs rattlers. In their analysis, the Cs cations are interpreted as damped Einstein oscillators with an effective frequency of 4 meV \cite{lahns24a}. Furthermore, phonon dispersion calculations of a system with artificially raised Cs masses demonstrate an increased interference of the Cs rattling with the acoustic phonon modes. Our interpretation, now supported by the MD calculations, is that the close proximity to the halide anions makes the Cs rattlers to become coherently coupled to matching vibrational modes of the inorganic cage involving octahedral tilting. The origin of the anomalous EP coupling are these coupled modes, which provide additional oscillator strength to the phonon DOS, represented by the additional Einstein oscillator. In contrast, at low Cl concentrations, where NCs crystallize in the cubic phase, the mean tilt angle is zero and the unfolded Cs dynamics is limited to the central region of the cage voids. This, together with the larger cage size, could be the cause of the loss of coherence, leading to the suppression of the anomalous EP coupling. We refer to Note $\sharp$7 of S.I. for clarification, where we also provide a tentative explanation for the negative sign of the anomalous EP term based on the Fr\"{o}hlich interaction \cite{fuxxx18a}.


\section{Conclusions}

In conclusion, we elucidated the reason for the sign reversal of the temperature slope of the gap, when comparing CsPbCl$_3$ NCs (negative slope) with their bromide counterparts (positive slope). For this purpose a series of CsPb(Br$_{1-x}$Cl$_x$)$_3$ NCs with five different Cl contents were synthesized by anionic exchange. Using temperature and pressure-dependent PL, we univocally disentangle the contributions from thermal expansion and electron-phonon interaction. The aforementioned slope sign turnover is triggered by the occurrence of a structural phase transition, taking place at a Cl concentration of ca. 40\%, and the effects this transition has on the nature of the EP coupling term.
For low Cl concentrations the NCs are cubic \cite{brenn19a} and exhibit around room temperature a linear gap temperature dependence with the typical positive slope of MAPbI$_3$ \cite{franc19a}. Here the EP term accounts for the "normal" coupling to the inorganic cage phonons. On the contrary, for orthorhombic NCs with high Cl concentrations the gap temperature slope is negative as for CsPbCl$_3$ NCs \cite{saran17a}. The only way to account for the sign reversal is to introduce an additional EP term with negative amplitude, which becomes the leading term in the gap renormalization. We ascribe it to an {\it anomalous} coupling to mixed modes arising from Cs rattler modes intermixed with cage vibrations involving octahedral tilting. Given the relevance of the gap temperature dependence for the optoelectronic properties of perovskite NCs, its correct assessment is key for the advancement of emergent photovoltaics and efficient light emission and/or sensing devices, for instance.


\section{Supporting Information} Contains details of the PL and Raman characterization of the entire series of CsPb(Br,Cl)$_3$ NCs, in particular the determination of the NC composition from the energy of the PL peak maximum. It also contains the cascade plots of the PL spectra and the gap-energy data points as a function of temperature and pressure for the samples with the Cl concentrations not shown in the manuscript. It also includes the smoothed energy-vs-temperature data points sets for all compositions and the first-derivative plots for the three Cl contents not shown in the main text. A table with the parameters resulting from the least-squares fits to the first derivative data points is also included. Finally, a detailed discussion of the underlying physics of the normal and anomalous EP term is provided.

\section{Acknowledgements}

The Spanish "Ministerio de Ciencia, Innovaci\'{o}n y Universidades" (MICIU) through the Agencia Estatal de Investigaci\'{o}n (AEI) is gratefully acknowledged for its support through grant CEX2023-001263-S (MATRANS42) in the framework of the Spanish Severo Ochoa Centre of Excellence program and the AEI/FEDER(UE) grants PID2020-117371RA-I00 (CHIRALPERO), PID2021-128924OB-I00 (ISOSCELLES), PID2022-141956NB-I00 (OUTLIGHT) and TED2021-131628A-I00 (MACLEDS). The authors also thank the Catalan agency AGAUR for grant 2021-SGR-00444 and the National Network "Red Perovskitas" (MICIU funded). S.F. acknowledges a FPI grant PRE2021-100097 from MICIU and the PhD programme in Materials Science from Universitat Aut\`{o}noma de Barcelona in which she is enrolled. B.S. acknowledges the support of the Erasmus+ programme of the European Union through the internship project 2022-1-DE01-KA131-HED-000055364 (STREAM 2022). L.P. acknowledges support from the Spanish MICIU through Ram\'{o}n y Cajal grant (RYC2018-026103-I) and a grant from the Xunta de Galicia (ED431F2021/05). S.G.G. acknowledges support from project CNS2022-135531 (HARDTOP) funded by 14 MCIN/AEI/10.13039/501100011033.



{\bf References}\\



\section*{Supplementary material}

Table S 2: Einstein-oscillator parameters corresponding to the effective amplitude $A_i$ and frequency $\omega_i$ describing the contribution from electron-phonon interaction $\left[\frac{\partial E_g}{\partial T}\right]_{EP}$ to the renormalization of the band gap in CsPb(Br$_{1-x}$Cl$_x$)$_3$ NCs. Numbers in parentheses are error bars (uncertainty of the last digits). Values marked with an asterisk are from CsPbBr$_3$ NCs.[1]

| x | Einstein osc. | $A_i$ (meV) | $\hbar\omega_i$ (meV) |
|---|---|---|---|
| 0* | Osc. ♯1 | 6.7(5)* | 6* |
| 0.10 | Osc. ♯1 | 4.5(5) | 6 |
| 0.25 | Osc. ♯1 | 12.5(5) | 6 |
| 0.40 | Osc. ♯1 | 7 | 6 |
|  | Osc. ♯2 | -13(2) | 4 |
| 0.75 | Osc. ♯1 | 7 | 6 |
|  | Osc. ♯2 | -20(1) | 4 |
| 0.85 | Osc. ♯1 | 7 | 6 |
|  | Osc. ♯2 | -20(1) | 4 |

Here, we would like to point out that the model with two Einstein oscillators has been previously used by Saran et al.[2] in an attempt to describe the temperature dependence of the gap of CsPbI$_3$ and CsPbBr$_3$ NCs in the temperature range from 5 to 300 K. The two oscillators were supposed to account for the coupling of electrons with acoustic and optical phonon modes of the perovskite, the latter mediated by Fröhlich interaction. Unfortunately, in this work two crucial issues were not taken into account. The first and most fundamental is that thermal expansion effects were ignored, probably misled by the results on cuprous halides[15] and silver chalcopyrites.[25] As previously pointed out by Francisco-López et al.,[26] discarding the TE term leads to an overestimation of the electron-phonon interaction, such that the outcome of its modeling in terms of Einstein

oscillators yielded unphysically large values for the frequencies of the involved phonons (larger than the cutoff frequency for the inorganic cage phonons which is roughly 25 meV). The second problem arises when the authors try to cover the temperature dependence of the gap over the entire range of the experiment for which the temperature dependence of the gap is outspoken non-linear. As deduced from the discussion in Note ♯3, such non-linearity arises to a large extent because the TE term is temperature dependent as well. Evidence of that are the different values obtained for the gap pressure coefficient of MAPbI$_3$ for different temperatures within the range of stability of the orthorhombic phase.[27] This is the reason for the need of two oscillators, i.e. to describe the nonlinearity, while we clearly show that, in the linearity regime around room temperature (constant TE term), the gap temperature renormalization of CsPbBr$_3$ NCs is perfectly described by a single oscillator for the EP term. Moreover, the electron-phonon coupling is in this case the *normal* one, as exhibited by most lead halide perovskites.[1]

## Note ♯6: Normal electron-phonon coupling

As mentioned before, the normal EP coupling term is described in terms of a single Einstein-oscillator model, which contains two parameters: The oscillator amplitude $A_{eff}$ and its characteristic frequency $\omega_{eff}$. In principle, the latter should correspond to a given peak in the phonon density of states (DOS). For that purpose we reproduce the calculations of the phonon dispersion curves and the corresponding DOS from Ref.[9] The vibrational frequencies for the cubic phase of MAPbBr$_3$ were calculated within the harmonic phonon approximation using second order force constants obtained from density functional theory (DFT). Figure S11 (left panel) shows the phonon dispersions of the 18 inorganic cage modes within the harmonic approximation using a pseudo cubic lattice for MAPbBr$_3$ at room temperature. Negative-frequency soft modes are located around the Brillouin-zone boundary at the M and R points. The phonon dispersions are plotted considering

band crossings. The color refers to the nature of each phonon eigenmode. The three orthogonal acoustic modes are plotted in blue shades. The remaining modes are optical, plotted in groups of three with a similar shade for each orthogonal mode (these modes would be degenerate if the MA ions were replaced by a spherical atom). Figure S11 (right panel) displays the corresponding phonon DOS decomposed by sets of three orthogonal phonon eigenmodes, integrated over the full Brillouin zone, but not considering band crossings.

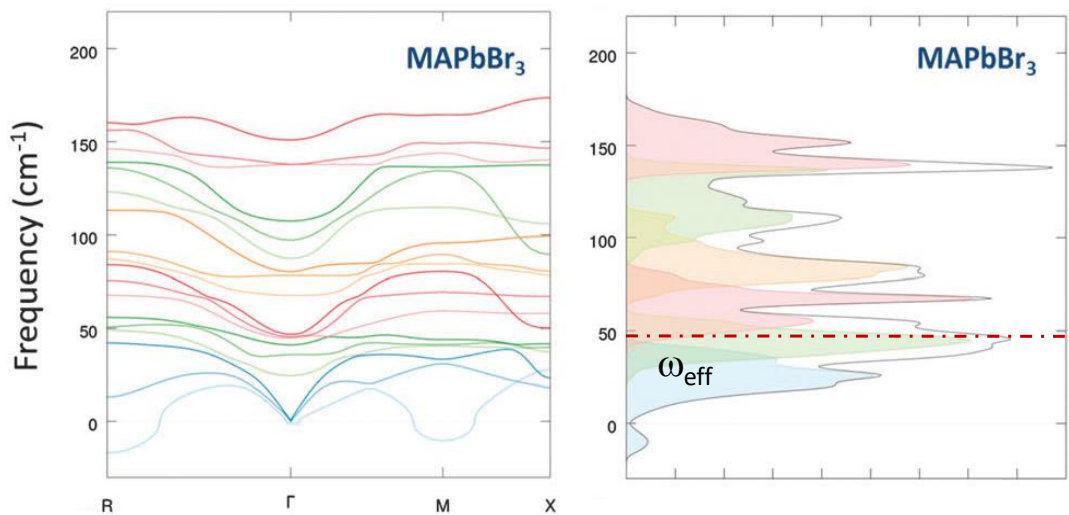

Figure S 11: (Left panel) Phonon dispersion curves and (Right panel) phonon density of states (DOS) of bulk MAPbBr$_3$ calculated within the harmonic phonon approximation using second order force constants obtained from density functional theory (adapted from Fig. 3 of Ref.[9]).

The dot-dashed horizontal line in the right panel of Fig. S11 represents the effective frequency of the Einstein oscillator used to describe the *normal* EP coupling term in the case of the CsPbBr$_3$ NCs. An inspection of the phonon DOS indicates that the oscillator frequency of 6 meV (i.e. ca. 48 cm$^{-1}$) lies slightly above the first well-defined peak or band in the DOS, corresponding to low-frequency optical modes partially intermixed with acoustical branches. Since these are all phonons of the inorganic cage, we can safely assume that a similar oscillator will account for the EP term

in the CsPbBr$_3$ NCs as well. This is so because both MAPbBr$_3$ and CsPbBr$_3$ share the same inorganic cage, being the influence of the A-site cation negligible to first order of approximation. Furthermore, we note that the contributions of the higher-frequency peaks in the phonon DOS are exponentially damped by the Bose-Einstein phonon occupation number. This explains why a single Einstein oscillator is enough to account for the coupling between the electrons and inorganic cage phonons.

## Note ♯7: Anomalous electron-phonon coupling

Equation (4) in Note ♯5 tells us two important things about the electron-phonon interaction: it is an additive magnitude, given by an integral over frequency of the electron-phonon spectral function $g^2F(n,\mathbf{k},\omega)$, and that there has to be a non-vanishing overlap between the electronic and vibrational wavefunctions for the matrix elements $\frac{\partial E_{n\mathbf{k}}}{\partial n_{j\mathbf{q}}}$ to be different from zero. Since the states near the valence and conduction band edges in lead halide perovskites are completely confined to the inorganic cage, the EP term only captures contributions from lattice phonons, excluding the A-site cations. This is at the origin of the *normal* EP coupling term, represented by the Einstein oscillator with positive amplitude. However, the A-site cations might also add oscillator strength into the electron-phonon spectral function by coupling indirectly with the charge carriers in the inorganic cage. A way in which this indirect coupling can proceed is when a translational, rotational or librational degree of freedom of the A-site cations enters into resonance with any vibrational mode of the inorganic cage lattice. In the case of Cs being the A-site cation, the modes participating in the anomalous (indirect) EP coupling are the Cs rattlers,[28] as explained in the main manuscript. Due to the Coulombic nature of the anion-cation attraction, the most likely phonon modes involved in the indirect coupling are the ones associated to octahedral tilting. Please notice that this implies the formation of modes featuring *coherence* between the A-site cation motion (Cs rattlers, in this

case) and the octahedral-tilting phonons. This leads at last to the *anomalous* EP term which is accounted for within our model by the second Einstein oscillator with negative amplitude. We point out that incoherent coupling can only affect the lifetime (even the frequency) of the inorganic cage phonons[10] but never add oscillator strength to the EP spectral function.

Further support for the idea that octahedral tilting is involved in the anomalous EP coupling comes from recent first-principles density functional theory calculations of the energetics of octahedral tilting in $CsPbI_3$ and $CsSnI_3$.[29] The constant-energy surfaces of the atomic potential energy, which corresponds to the sum of the electronic kinetic energy and the potential energies due to electron-electron, electron-nucleus and nucleus-nucleus interactions,[30] were calculated as a function of octahedral tilting angle for both the case of in-phase and out-of-phase tilting between adjacent octahedrons.[29] In the orthorhombic *Pnma* phase, these potential energy surfaces exhibit multi-well landscape with four well-defined local minima separated by relatively small activation barriers, when tilting is only considered in a plane. At high enough temperatures, the atomic wave function delocalizes between the equivalent minima, leading to a dynamic tilt situation, which is more or less fast depending on the height of the barriers.[29,30] In contrast, at low temperatures the atoms localize at one of the four equivalent minima, which corresponds to a situation of static (fixed) octahedral tilting. In the work of Pérez-Fidalgo et al.,[31] it was proposed that the anomalous EP term needed to describe the reduction in temperature slope of the gap of low-content $Cs_xMA_{1-x}PbI_3$, occurring at around ambient temperature, arises from dynamic tilting fluctuations of the $PbI_6^-$ octahedrons in synchrony with the motion of the $Cs^+$ cations. This was observed at temperatures higher than the onset temperature for unfolding the $Cs^+$ cation dynamics.

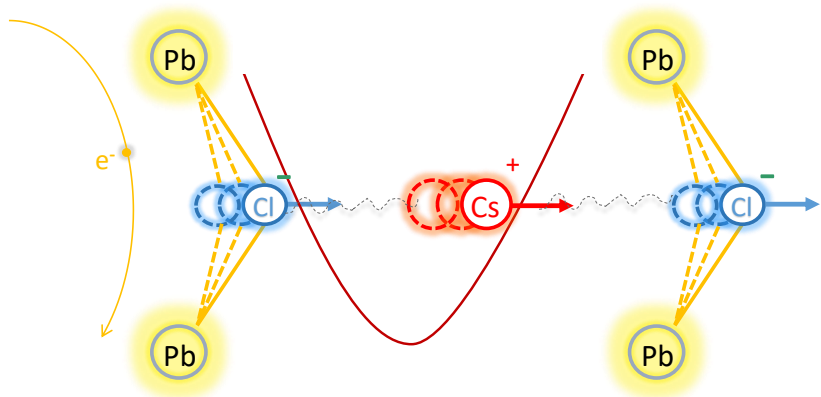

Figure S 12: Sketch illustrating the anomalous EP term arising from the (in-phase) coherent coupling of phonon modes of the inorganic cage involving octahedral tilting and the oscillations of the Cs cations in the potential energy well within the cage voids.

In contrast, a different situation is observed for $CsPb(Br,Cl)_3$ NCs, because the anomalous EP coupling appears for the shrunken orthorhombic phase at high Cl concentrations, for which the octahedral tilting is static and the dynamics of the $Cs^+$ cations is frozen. Nevertheless, we rationalize this finding as due to the reduced steric volume left for the $Cs^+$ cations inside the cage voids of the orthorhombic phase. As illustrated in the sketch of Fig. S12, coherence between the vibrations involving tilting and Cs oscillations in the potential well of the void seems to be established via the attractive anion-cation interaction. On the contrary, for the cubic Br-rich NCs the constant energy surfaces exhibit a single potential well centered at zero tilt, as is mandatory for a cubic phase. Consequently, although the dynamics of the Cs cation is unfolded, the motion is restricted to the central region of the cage voids and is characterized by a spherical atomic probability density cloud. This, combined with the larger cage size, could be the cause of the loss of coherence, so only the normal EP coupling remains; the one involving only the phonons of the inorganic cage.

In order to visualize the rattler modes involved in the anomalous, indirect EP coupling, we reproduce in Fig. S13 the results for the harmonic phonon dispersion relations of cubic CsPbBr$_3$ from large-scale molecular dynamics (MD) simulations combined with machine-learning interatomic potentials.[28] By highlighting in red color the Cs contribution in the respective eigenmode, one can identify nearly dispersionless rattling bands. In the middle panel, one can see that the Cs cations mainly contribute to the phonon band structure in an energy window around ca. 1 THz ($\approx$ 4 mev). This frequency perfectly matches that of acoustic phonons of the inorganic cage, which exhibit *in-phase* octahedral tilting.[9] In the left(right) panel, the masses of only the Cs atoms were artificially lowered(raised), which results in a blue(red)-shift of the central rattling frequency, respectively. For the light Cs cations, the central frequency shifts up to about 2.2 THz, thereby almost removing the overlap with the linear acoustic bands.

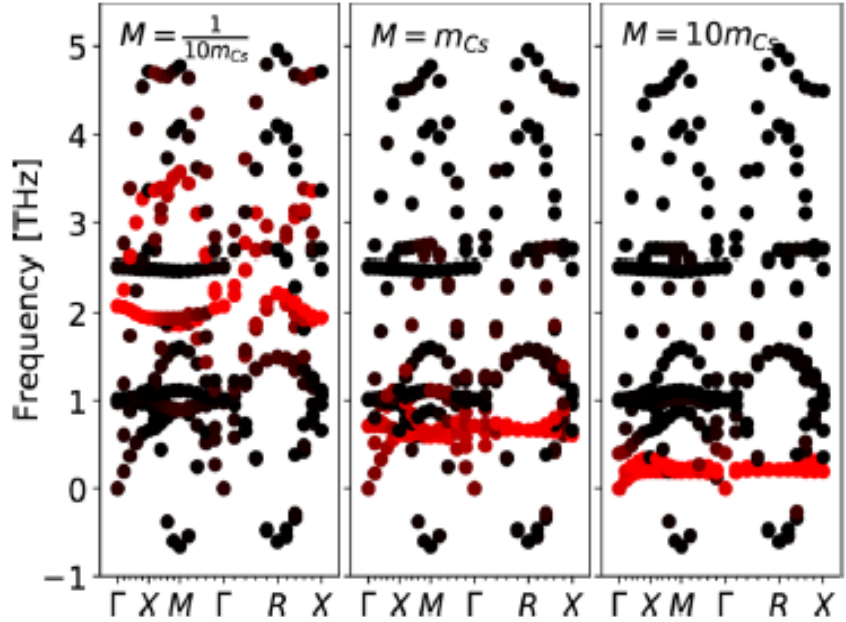

Figure S 13: Shifting the rattling band. Harmonic phonon band structure of cubic CsPbBr$_3$ with light ($m_{Cs}/10$), normal ($m_{Cs}$), and heavy ($m_{Cs} \cdot 10$) effective mass of the Cs cation. The color scale from black to red indicates the increasing Cs contribution to the respective eigenmode.




\begin{thebibliography}{99}

\bibitem{ahmed20a} Ahmed, G. H.; Yin, J.; Bakr, O. M.; Mohammed, O. F. Near-Unity Photoluminescence Quantum Yield in Inorganic Perovskite Nanocrystals by Metal-Ion Doping. {\it J. Chem. Phys.} {\bf 2020}, 152, 020902/1-10.
\bibitem{meixx22a} Mei, X.; Zhang, L.; Zhang, X.; Ding, L. Perovskite Nanocrystals for Light-Emitting Diodes. {\it J. Semicond.} {\bf 2022}, 43, 090201/1-5.
\bibitem{romer23a} Romero-P\'{e}rez, C.; Fern\'{a}ndez Delgado, N.; Herrera-Collado, M.; Calvo, M. E.; M\'{\i}guez, H. Ultrapure Green High Photoluminescence Quantum Yield from FAPbBr$_3$ Nanocrystals Embedded in Transparent Porous Films. {\it Chem. Mater.} {\bf 2023}, 35, 5541-5549.
\bibitem{shixx17a} Shi, Z.-F.; Li, Y.; Li, S.; Ji, H.-F.; Lei, L.-Z.; Wu, D.; Xu, T.-T.; Xu, J.-M.; Tian, Y.-T.; Li, X.-J. Polarized Emission Effect Realized in CH$_3$NH$_3$PbI$_3$ Perovskite Nanocrystals. {\it J. Mater. Chem. C} {\bf 2017}, 5, 8699-8706.
\bibitem{rubin21a} Rubino, A.; Francisco-L\'{o}pez, A.; Barker, A. J.; Petrozza, A.; Calvo, M. E.; Go\~{n}i, A. R.; M\'{\i}guez, H. Disentangling Electron-Phonon Coupling and Thermal Expansion Effects in the Band Gap Renormalization of Perovskite Nanocrystals. {\it J. Phys. Chem. Lett.} {\bf 2021}, 12, 569-575.
\bibitem{fangx17a} Fang, H.-H.; Protesescu, L.; Balazs, D. M.; Adjokatse, S.; Kovalenko, M. V.; Loi, M. A. Exciton Recombination in Formamidinium Lead Triiodide: Nanocrystals versus Thin Films. {\it Small} {\bf 2017}, 13, 1700673/1-10.
\bibitem{dirol18a} Diroll, B. T.; Guo, P.; Schaller, R. D. Unique Optical Properties of Methylammonium Lead Iodide Nanocrystals Below the Bulk Tetragonal-Orthorhombic Phase Transition. {\it Nano Lett.} {\bf 2018}, 18, 846-852.
\bibitem{wooxx18a} Woo, H. C.; Choi, J. W.; Shin, J.; Chin, S.-H.; Ann, M. H.; Lee, C.-L. Temperature-Dependent Photoluminescence of CH$_3$NH$_3$PbBr$_3$ Perovskite Quantum Dots and Bulk Counterparts. {\it J. Phys. Chem. Lett.} {\bf 2018}, 9, 4066-4074.
\bibitem{sadhu19a} Sadhukhan, P.; Pradhan, A.; Mukherjee, S.; Sengupta, P.; Roy, A.; Bhunia, S.; Das, S. Low Temperature Excitonic Spectroscopy Study of Mechano-Synthesized Hybrid Perovskite. {\it Appl. Phys. Lett.} {\bf 2019}, 114, 131102/1-5.
\bibitem{liuxx19a} Liu, L.; Zhao, R.; Xiao, C.; Zhang, F.; Pevere, F.; Shi, K.; Huang, H.; Zhong, H.; Sychugov, I. Size-Dependent Phase Transition in Perovskite Nanocrystals. {\it J. Phys. Chem. Lett.} {\bf 2019}, 10, 5451-5457.
\bibitem{lixxx22a} Li, J.; Guo, Z.; Xiao, S.; Tu, Y.; He, T.; Zhang, W. Optimizing Optical Properties of Hybrid Core/Shell Perovskite Nanocrystals. {\it Inorg. Chem. Front.} {\bf 2022}, 9, 2980-2986.
\bibitem{saran17a} Saran, R.; Heuer-Jungemann, A.; Kanaras, A. G.; Curry, R. J. Giant Bandgap Renormalization and Exciton-Phonon Scattering in Perovskite Nanocrystals. {\it Adv. Optical Mater.} {\bf 2017}, 5, 1700231/1-9.
\bibitem{leexx17a} Lee, S. M.; Moon, C. J.; Lim, H.; Lee, Y.; Choi, M. Y.; Bang, J. Temperature-Dependent Photoluminescence of Cesium Lead Halide Perovskite Quantum Dots: Splitting of the Photoluminescence Peaks of CsPbBr$_3$ and CsPb(Br/I)$_3$ Quantum Dots at Low Temperature. {\it J. Phys. Chem. C} {\bf 2017}, 121, 26054-26062.
\bibitem{gauxx23a} Gau, D. L.; Galain, I.; Aguiar, I.; Marotti, R. E. Origin of Photoluminescence and Experimental Determination of Exciton Binding Energy, Exciton-Phonon Interaction, and Urbach Energy in $\gamma$-CsPbI$_3$ Nanoparticles. {\it J. Lumin.} {\bf 2023}, 257, 119765/1-9.
\bibitem{lixxx16a} Li, J.; Yuan, X.; Jing, P.; Li, J.; Wei, M.; Hua, J.; Zhao, J.; Tian, L. Temperature-Dependent Photoluminescence of Inorganic Perovskite Nanocrystal Films. {\it RSC Adv.} {\bf 2016}, 6, 78311-78316.
\bibitem{shind17a} Shinde, A.; Gahlaut, R.; Mahamuni, S. Low-Temperature Photoluminescence Studies of CsPbBr$_3$ Quantum Dots. {\it J. Phys. Chem. C}  {\bf 2017}, 121, 14872-14878.
\bibitem{zhang19a} Zhang, X.; Gao, X.; Pang, G.; He, T.; Xing, G.; Chen, R. Effects of Material Dimensionality on the Optical Properties of CsPbBr$_3$ Nanomaterials. {\it J. Phys. Chem. C} {\bf 2019}, 123, 28893-28897.
\bibitem{strand21a} Strandell, D. P.; Kambhampati, P. The Temperature Dependence of the Photoluminescence of CsPbBr$_3$ Nanocrystals Reveals Phase Transitions and Homogeneous Linewidths. {\it J. Phys. Chem. C} {\bf 2021}, 125, 27504-27508.
\bibitem{xuxxx23a} Xu, F.; Wei, H.; Wu, Y.; Zhou, Y.; Li, J.; Cao, B. Nonmonotonic Temperature-Dependent Bandgap Change of CsPbCl$_3$ Films Induced by Optical Phonon Scattering. {\it J. Lumin.} {\bf 2023}, 257, 119736/1-7.
\bibitem{huang22a} Huang, X.; Li, X.; Tao, Y.; Guo, S.; Gu, J.; Hong, H.; Yao, Y.; Guan, Y.; Gao, Y.; Li, C.; Lu, X.; Fu, Y. Understanding Electron-Phonon Interactions in 3D Lead Halide Perovskites from the Stereochemical Expression of 6s$^2$ Lone Pairs. {\it J. Am. Chem. Soc.} {\bf 2022}, 144, 12247-12260.
\bibitem{golds26a} Goldschmidt, V. M. Die Gesetze der Krystallochemie, {\it Die Naturwissenschaften} {\bf 1926}, 14, 477-485.
\bibitem{patri15a} Patrick, C. E.; Jacobsen, K. W.; Thygesen, K. S. Anharmonic Stabilization and Band Gap Renormalization in the Perovskite CsSnI$_3$. {\it Phys. Rev. B} {\bf 2015}, 92, 201205R/1-5.
\bibitem{huang21a} Huang, Y.; Zhou, J.; Peng, L.; Li, K.; Elliott, S. R.; Sun, Z. Antibonding-Induced Anomalous Temperature Dependence of the Band Gap in Crystalline Ge$_2$Sb$_2$Te$_5$. {\it J. Phys. Chem. C} {\bf 2021}, 125, 19537-19543.
\bibitem{singh22a} Singh, A.; Satapathi, S. Phonon-Assisted Reversible Thermochromism in a Lead-Free Antimony-Based Cs$_3$Sb$_2$Br$_9$ Perovskite. {\it ACS Appl. Electron. Mater.} {\bf 2022}, 4, 3440-3447.
\bibitem{laute85a} Lautenschlager, P.; Allen, P. B.; Cardona, M. Temperature Dependence of Band Gaps in Si and Ge. {\it Phys. Rev. B} {\bf 1985}, 31, 2163-2171.
\bibitem{gopal87a} Gopalan, S.; Lautenschlager, P.; Cardona, M. Temperature Dependence of the Shifts and Broadenings of the Critical Points in GaAs. {\it Phys. Rev. B} {\bf 1987}, 35, 5577-5584.
\bibitem{laute87a} Lautenschlager, P.; Garriga, M.; Logothetidis, S.; Cardona, M. Interband Critical Points of GaAs and their Temperature Dependence. {\it Phys. Rev. B} {\bf 1987}, 35, 9174-9189.
\bibitem{cardo89a} Cardona, M.; Gopalan, S. Temperature Dependence of the Band Structure of Semiconductors: Electron-Phonon Interaction. {\it Progress on Electron Properties of Solids}, ed. Girlanda, R.; et al. (Kluver, 1989), p. 51-64.
\bibitem{shang23a} Shang, H.; Yang, J. The Electron-Phonon Renormalization in the Electronic Structure Calculation: Fundamentals, Current Status, and Challenges. {\it J. Chem. Phys.} {\bf 2023}, 158, 130901/1-11.
\bibitem{franc19a} Francisco L\'{o}pez, A.; Charles, B.; Weber, O. J.; Alonso, M. I.; Garriga, M.; Campoy-Quiles, M.; Weller, M. T.; Go\~{n}i, A. R. Equal Footing of Thermal Expansion and Electron-Phonon Interaction in the Temperature Dependence of Lead Halide Perovskite Band Gaps. {\it J. Phys. Chem. Lett.} {\bf 2019}, 10, 2971-2977.
\bibitem{franc18a} Francisco L\'{o}pez, A.; Charles, B.; Weber, O. J.; Alonso, M. I.; Garriga, M.; Campoy-Quiles, M.; Weller, M. T.; Go\~{n}i, A. R. Pressure-Induced Locking of Methylammonium Cations Versus Amorphization in Hybrid Lead Iodide Perovskites. {\it J. Phys. Chem. C} {\bf 2018}, 122, 22073-22082.
\bibitem{perez23a} P\'{e}rez-Fidalgo, L.; Xu, K.; Charles, B. L.; Henry, P. F.; Weller, M. T.; Alonso, M. I.; Go\~{n}i, A. R. Anomalous Electron-Phonon Coupling in Cesium-Substituted Methylammonium Lead Iodide Perovskites. {\it J. Phys. Chem. C} {\bf 2023}, 127, 22817-22826.
\bibitem{lahns24a} Lahnsteiner, J.; Rang, M.; Bokdam, M. Tuning Einstein Oscillator Frequencies of Cation Rattlers: A Molecular Dynamics Study of the Lattice Thermal Conductivity of CsPbBr$_3$. {\it J. Phys. Chem. C} {\bf 2024}, 128, 1341-1349.
\bibitem{tongx16a} Tong, Y.; Bladt, E.; Ayg\"{u}ler, M. F.; Manzi, A.; Milowska, K. Z.; Hintermayr, V. A.; Docampo, P.; Bals, S.; Urban, A. S.; Polavarapu, L.; Feldmann, J. Highly Luminescent Cesium Lead Halide Perovskite Nanocrystals with Tunable Composition and Thickness by Ultrasonication. {\it Angew. Chem. Int. Ed.} {\bf 2016}, 55, 13887-13892.
\bibitem{fasah24a} Fasahat, S.; Sch\"{a}fer, B.; Xu, K.; Fiuza-Maneiro, N.; G\'{o}mez-Gra\~{n}a, S.; Alonso, M. I.; Polavarapu, L.; Go\~{n}i, A. R. Absence of Anomalous Electron-Phonon Coupling in the Temperature Renormalization of the Gap of CsPbBr$_3$ Nanocrystals. arXiv:2409.06374 [cond-mat.mtrl-sci] (https://doi.org/10.48550/arXiv.2409.06374). 
\bibitem{prote15a} Protesescu, L.; Yakunin, S.; Bodnarchuk, M. I.; Krieg, F.; Caputo, R.; Hendon, C. H.; Yang, R. X.; Walsh, A.; Kovalenko, M. V. Nanocrystals of Cesium Lead Halide Perovskites (CsPbX$_3$, X = Cl, Br, and I): Novel Optoelectronic Materials Showing Bright Emission with Wide Color Gamut. {\it Nano Lett.} {\bf 2015}, 15, 3692-3696.
\bibitem{hoffm23a} Hoffman, A. E. J.; Saha, R. A.; Borgmans, S.; Puech, P.; Braeckevelt, T.; Roeffaers, M. B. J.; Steele, J. A.; Hofkens, J.; Van Speybroeck, V. Understanding the Phase Transition Mechanism in the Lead Halide Perovskite CsPbBr$_3$ via Theoretical and Experimental GIWAXS and Raman Spectroscopy. {\it APL Mater.} {\bf 2023}, 11, 041124/1-12.
\bibitem{manni20a} Mannino, G.; Deretzis, I.; Smecca, E.; La Magna, A.; Alberti, A.; Ceratti, D.; Cahen, D. Temperature-Dependent Optical Band Gap in CsPbBr$_3$, MAPbBr$_3$, and FAPbBr$_3$ Single Crystals. {\it J. Phys. Chem. Lett.} {\bf 2020}, 11, 2490-2496.
\bibitem{brenn19a} Brennan, M. C.; Kuno, M.; Rouvimov, S. Crystal Structure of Individual CsPbBr$_3$ Perovskite Nanocubes. {\it Inorg. Chem.} {\bf 2019}, 58, 1555-1560.
\bibitem{xuxxx23b} Xu, K.; P\'{e}rez-Fidalgo, L.; Charles, B. L.; Weller, M. T.; Alonso, M. I.; Go\~{n}i, A. R. Using Pressure to Unravel the Structure\textemdash Dynamic-Disorder Relationship in Metal Halide Perovskites. {\it Sci. Rep.} {\bf 2023}, 13, 9300/1-12.
\bibitem{gonix24a} Go\~{n}i, A. R. Raman Linewidths as a Probe of Lattice Anharmonicity and Dynamic Disorder in Metal Halide Perovskites. {\it Asian J. Phys.} {\bf 2024}, 33, 29-38.
\bibitem{hanse24a} Hansen, K. R.; Colton, J. S.; Whittaker-Brooks, L. Measuring the Exciton Binding Energy: Learning from a Decade of Measurements on Halide Perovskites and Transition Metal Dichalcogenides. {\it Adv. Optical Mater.} {\bf 2024}, 12, 2301659/1-136.
\bibitem{goebe98a} G\"{o}bel, A.; Ruf, T.; Cardona, M.; Lin, C. T.; Wrzesinski, J.; Steube, M.; Reimann, K.; Merle, J.-C.; Joucla, M. Effects of the Isotopic Composition on the Fundamental Gap of CuCl. {\it Phys. Rev. B} {\bf 1998}, 57, 15183-15190.
\bibitem{nitsc02a} Nitsch, K.; Rodov\'{a}, M. Thermomechanical Measurements of Lead Halide Single Crystals. {\it phys. stat. sol. (b)} {\bf 2002}, 234, 701-709.
\bibitem{ezzel21a} Ezzeldien, M.; Al-Qaisi, S.; Alrowaili, Z. A.; Alzaid, M.; Maskar, E.; Es-Smairi, A.; Vu, T. V.; Rai, D. P. Electronic and Optical Properties of Bulk and Surface of CsPbBr$_3$ Inorganic Halide Perovskite a First Principles DFT 1/2 Approach. {\it Sci. Rep.} {\bf 2021} 11, 20622/1-12.
\bibitem{serra02a} Serrano, J.; Schweitzer, Ch.; Lin, C. T.; Reimann, K.; Cardona, M.; Fr\"{o}hlich, D. Electron-Phonon Renormalization of the Absorption Edge of the Cuprous Halides. {\it Phys. Rev. B} {\bf 2002}, 65, 125110/1-7.
\bibitem{bhosa12a} Bhosale, J.; Ramdas, A. K.; Burger, A.; Mu\~{n}oz, A.; Romero, A. H.; Cardona, M.; Lauck, R.; Kremer, R. K. Temperature Dependence of Band Gaps in Semiconductors: Electron-Phonon Interaction. {\it Phys. Rev. B} {\bf 2012}, 86, 195208/1-10.
\bibitem{leguy16a} Leguy, A. M. A.; Go\~{n}i, A. R.; Frost, J. M.; Skelton, J.; Brivio, F.; Rodr\'{\i}guez-Mart\'{\i}nez, X.; Weber, O. J.; Pallipurath, A.; Alonso, M. I.; Campoy-Quiles, M.; Weller, M. T.; Nelson, J.; Walsh, A.; Barnes, P. R. F. Dynamic Disorder, Phonon Lifetimes, and the Assignment of Modes to the Vibrational Spectra of Methylammonium Lead Halide Perovskites. {\it Phys. Chem. Chem. Phys.} {\bf 2016}, 18, 27051-27066.
\bibitem{brivi15a} Brivio, F.; Frost, J. M.; Skelton, J. M.; Jackson, A. J.; Weber, O. J.; Weller, M. T.; Go\~{n}i, A. R.; Leguy, A. M. A.; Barnes, P. R. F.; Walsh, A. Lattice Dynamics and Vibrational Spectra of the Orthorhombic, Tetragonal, and Cubic Phases of Methylammonium Lead Iodide. {\it Phys. Rev. B} {\bf 2015}, 92, 144308/1-8.
\bibitem{yaffe17a} Yaffe, O.; Guo, Y.; Tan, L. Z.; Egger, D. A.; Hull, T.; Stoumpos, C. C.; Zheng, F.; Heinz, T. F.; Kronik, L.; Kanatzidis, M. G.; et al. Local Polar Fluctuations in Lead Halide Perovskite Crystals. {\it Phys. Rev. Lett.} {\bf 2017}, 118, 136001.
\bibitem{klarb19a} Klarbring, J. Low-Energy Paths for Octahedral Tilting in Inorganic Halide Perovskites. {\it Phys. Rev. B} {\bf 2019}, 99, 104105/1-7.
\bibitem{adams23a} Adams, D. J.; Churakov, S. V. Classification of Perovskite Structural Types with Dynamical Octahedral Tilting. {\it IUCrJ} {\bf 2023}, 10, 1-12.
\bibitem{fuxxx18a} Fu, M.; Tamarat, P.; Trebbia, J.-B.; Bodnarchuk, M. I.; Kovalenko, M. V.; Even, J.; Lounis, B. Unraveling Exciton-Phonon Coupling in Individual FAPbI$_3$ Nanocrystals Emitting Near-Infrared Single Photons. {\it Nat. Commun.} {\bf 2018}, 9, 3318/1-10.

\end{thebibliography}
\end{document}